\documentclass[letter]{acoust}
\usepackage{fancybox}

\def\header{Making generalized voice morphing accessible} 

\title{Interactive tools for making temporally variable, multiple-attributes, and multiple-instances morphing accessible}
\subtitle{Flexible manipulation of divergent speech instances for explorational research and education}

 \author{
Hideki Kawahara$^1$,  and Masanori Morise$^2$\thanks{kawahara@wakayama-u.ac.jp, and mmorise@meiji.ac.jp}}
\inst{
$^{1}$ Center for Innovative and Joint Research, Wakayama University,  930 Sakaedani, Wakayama, Wakayama 640-8510, Japan\\
$^{2}$ Graduate School of Advanced Mathematical Sciences, Meiji University, 4-21-1, Nakano, Tokyo, 164-8525, Japan}

\abst{We generalized a voice morphing algorithm capable of handling temporally variable, multiple-attributes, and multiple instances.
The generalized morphing provides a new strategy for investigating speech diversity.
However, excessive complexity and the difficulty of preparation have prevented researchers and students from enjoying its benefits.
To address this issue, we introduced a set of interactive tools to make preparation and tests less cumbersome.
These tools are integrated into our previously reported interactive tools as extensions.
The introduction of the extended tools in lessons in graduate education was successful.
Finally, we outline further extensions to explore excessively complex morphing parameter settings.
}

\kword{Vocoder parameter, graphical user interface, interaction, generalized morphing, calicature}

\mladdress{Hideki Kawahara, 2-14-4 Kitayamato, Ikoma, Nara, 630-0121, Japan} 

\class{5} 

\recommend{2}  

\begin{document}
\maketitle

\section{Introduction}
Driven by the recent resurgence of deep learning, synthesized speech is already comparable to speech sounds spoken by humans~\cite{Tan2024ieeePami,cooper2024ast}.
Deep learning technologies also allow for converting voice identity and speaking style without introducing severe deterioration in their naturalness and quality~\cite{sisman2020ieee,huang2023asru,cooper2024ast}.
Synthesized speech and converted speech sounds based on those technologies seem now divergent enough to cover the divergence of speech by humans.
However, controlling these divergent speech sounds using physiologically and perceptually grounded relevant attributes is still challenging~\cite{sisman2020ieee,Tan2024ieeePami,huang2023asru,cooper2024ast}.

 We introduced interactive GUI tools 
 hoping to help users acquire a basic understanding of these relations~\cite{worldTools2023}.
We implemented the tools using a classical non-neural VOCODER called WORLD~\cite{morise2016world}.
We designed the tools to be upper compatible with the widely used classical VOCODER STRAIGHT~\cite{kawahara1999spcom,kawahara2008icassp} because a considerable amount of publications has been using STRAIGHT and STRAIGHT-based applications~\cite{smith2005jasa,zen2005ieicej,tokuda2013ieee,uchida2019jst,sisman2020ieee,Uchida2022ast}.

In this letter, we extend our WORLD-based tools~\cite{worldTools2023}, introducing generalized voice morphing by migrating STRAIGHT-based tools to the WORLD-based foundations.
The next section introduces voice morphing and its generalization, followed by the section describing the new morphing tools based on WORLD~\cite{morise2016world}.

\section{Voice morphing and generalization}
Voice morphing is a technique to generate intermediate (or exaggerated) voices made from a set of different voice instances.
Even the early development stage of the morphing procedure~\cite{kawahara2003icassp} has led to various new findings~\cite{Schweinberger2008currentbiol,Bruckert2010currentbiol}.
We introduced a generalized morphing procedure capable of morphing many instances by setting time-varying morphing rates to individual morphing parameters~\cite{kawahara2013apsipa,kawahara2015book}.
This generalization further produced many interesting findings~\cite{Skuk2014JSLHR,kawahara2019voice,Skuk2020JSLHR,Nussbaum2022cognition}.
However, its excessive complexity in preparation and handling prevented many researchers from using it for their investigations.
We extended our tools~\cite{worldTools2023} to make this generalized morphing more accessible to prospective users.

\subsection{Generalized morphing}
Classical VOCODERs decompose a speech signal into a set of VOCODER parameters~\cite{kawahara1999spcom,kawahara2008icassp,worldTools2023}.
They are filter and excitation source parameters.
The filter parameter is a time-indexed sequence of the power spectrum envelope.
The source parameters are a time-indexed sequence of aperiodicity and the fundamental frequency.

Generalized morphing interpolates and extrapolates these parameters and generates morphed speech sounds from the manipulated parameters.
Because these parameters are located on time and frequency axes, we added them to the parameters to be modified.
They are:
%
(1) $\mathbf{tx:}$
time axis, a function from index to time,
(2) $\mathbf{fx:}$
frequency axis, a function from index to frequency,
(3) $\mathbf{sl:}$
spectrum level, the power spectrum envelope value at a time-frequency location,
(4) $\mathbf{fo:}$
fundamental frequency at an indexed time location, and
(5) $\mathbf{ap:}$
aperiodicity, a power ratio of the random component and periodic component at a time-frequency location.

Let $\mathbf{x}$ represent one of the above parameters.
The following equation represents the parameter manipulation of generalized morphing~\cite{kawahara2013apsipa,kawahara2015book}.
\begin{align}
\mathbf{x}_\mathrm{morph} & = \mathcal{T}^{-1}\!\left[\sum_{k=1}^{K} w_k\mathcal{T}[ \mathbf{x}_k]\right] , \ \mbox{where} \ 
\sum_{k=1}^{K} w_k = 1 ,
\end{align}
where the function $\mathcal{T}[\mathbf{x}]$ maps the original restricted range of $\mathbf{x}$ onto the whole real number $\mathbb{R}$, and its inverse function $\mathcal{T}^{-1}$ maps the manipulated parameter back to the original restricted range.
The variable $w_k$ represents a real-valued weight 
to implement linear interpolation and extrapolation.

\subsection{STRAIGHT-based morphing tools}
STRAIGHT-based morphing tools implement functions $\mathbf{tx}$ and $\mathbf{fx}$ using piece-wise linear functions by assigning anchoring points~\cite{kawahara2013apsipa,kawahara2015book}.
We implemented an anchor assignment GUI tool and alignment tool using the \texttt{GUID}, an old UI design environment of \texttt{MATLAB}~\cite{MATLAB}.

First, the user selects one speech instance as the reference and assigns temporal anchoring points to the boundaries of acoustically different regions.
Then, the user assigns frequency anchoring points at prominent spectral peaks at each temporal anchor's location.

Second, the user selects the other speech instance as the target.
The alignment tool shows spectrographic representations of the reference and the target side by side.
The goal of alignment is to minimize the deterioration of the perceived quality of the morphed sounds. 
This time-consuming and cumbersome assignment and alignment procedures require a basic understanding of acoustic phonetics and digital signal processing.

\subsection{STRAIGHT-based morphing application}
For psycho-acoustics, we introduced the stimuli continuum generation tool.
For demonstrations, we developed interactive three-instance morphing~\cite{kawahara2015book}.

\section{New WORLD-based morphing tools}
The positive experience of using WORLD-based tools in class lessons~\cite{worldTools2023} encouraged us to expand their utility by introducing generalized morphing tools.
The significant performance boost of computation~\cite{Leiserson2020science} 
also motivated us. 
We aim to minimize the required prerequisite knowledge and make generalized morphing accessible.

\subsection{Assignment and alignment tool}
\begin{figure}[tbp]
\begin{center}
\includegraphics[width=0.8\hsize]{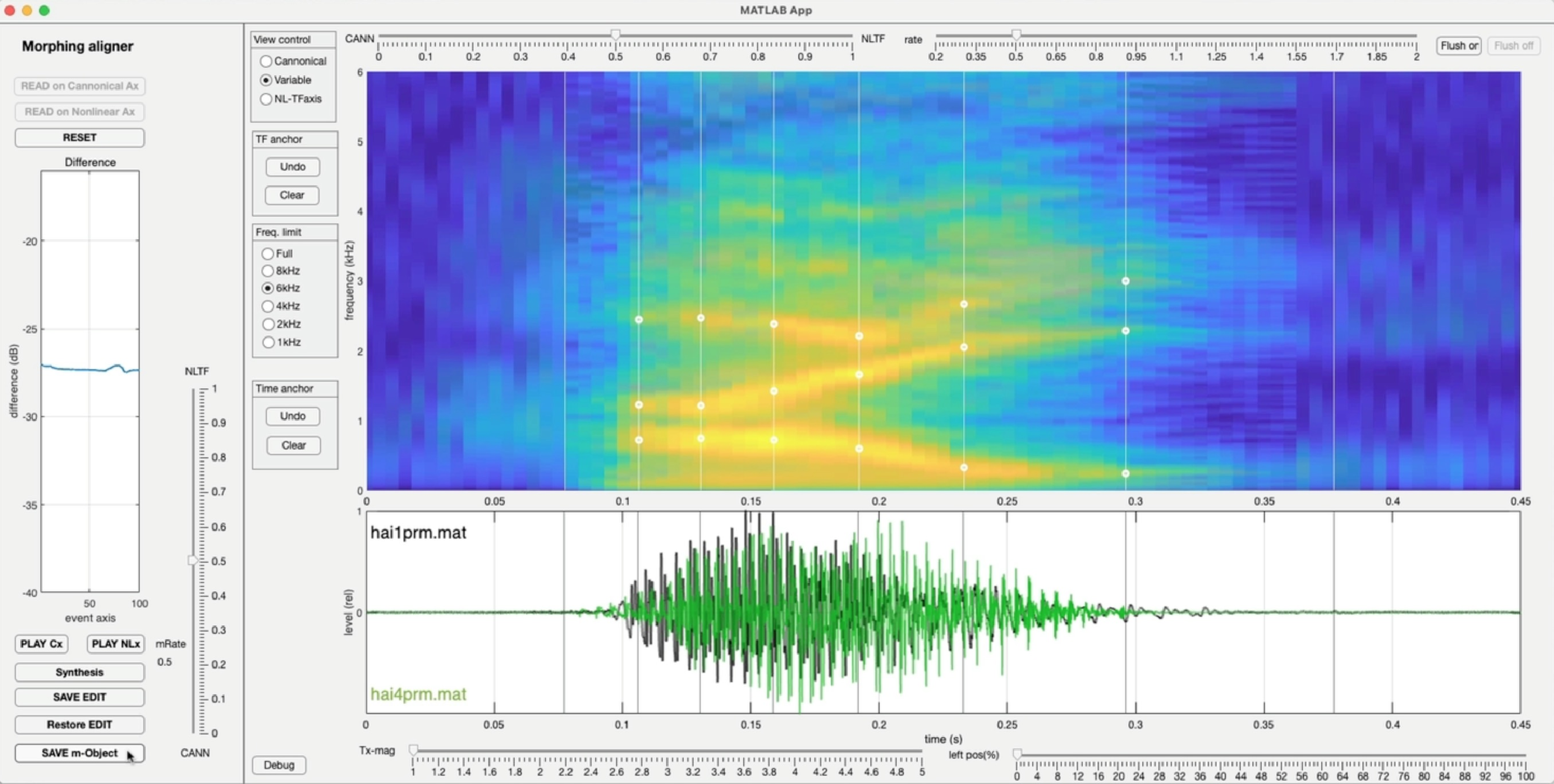}\\
\caption{GUI snapshot of the morphing object preparation. This snapshot shows the final stage of time-frequency alignment (at 6:41 of the video ``10'' in Fig.~\ref{fig:tutorialVideos}). The sample (WAVE: 44100~Hz, 32~bit) is a Japanese word /hai/ (``Yes'' in English) spoken by the first author.}
\label{fig:mObjectPreparationGUI}
\end{center}
\end{figure}

\begin{figure}[tbp]
\begin{center}
\includegraphics[width=0.75\hsize]{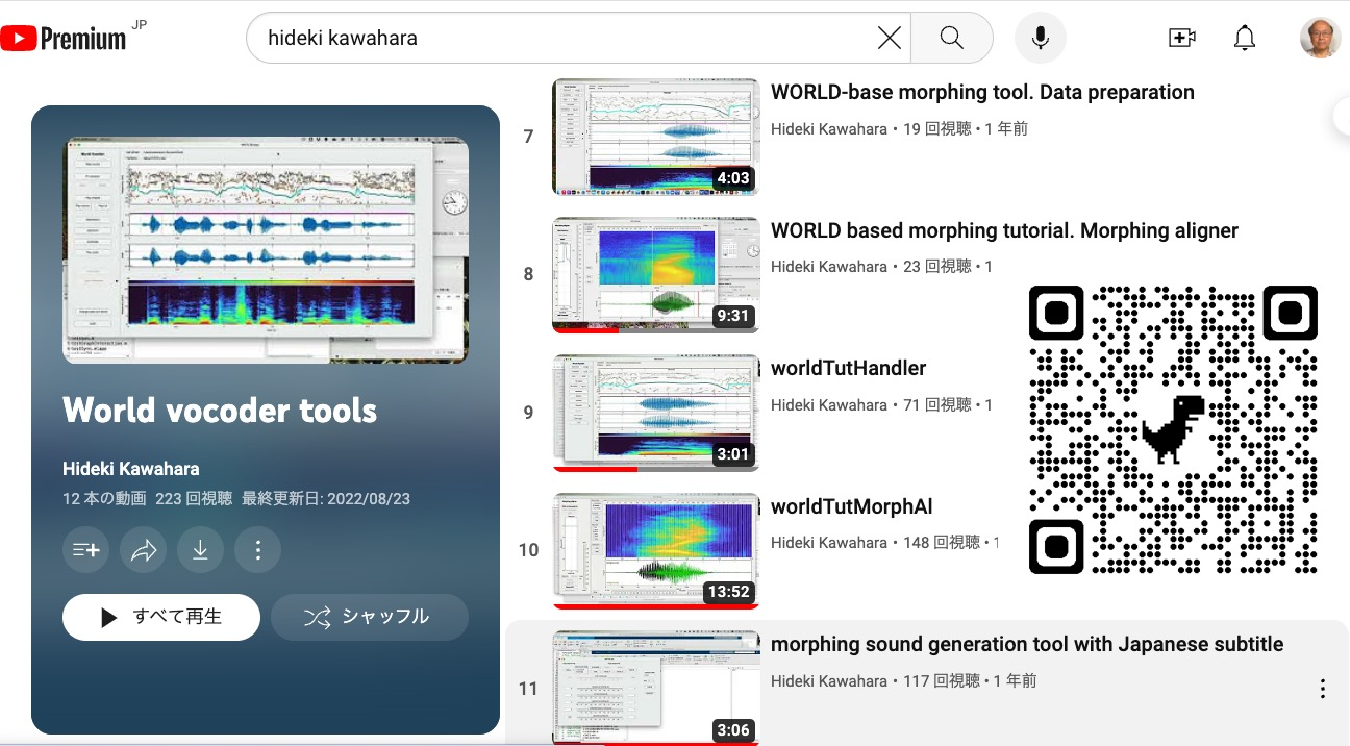}\\
\caption{Tutorial video channel of WORLD vocoder tools. This snapshot lists morphing tutorial videos. The QR code links to this YouTube channel~\cite{worldYouTube}.}
\label{fig:tutorialVideos}
\end{center}
\end{figure}

We integrated the anchor assignment and instance alignment tools and developed one GUI tool for morphing object preparation.
Figure~\ref{fig:mObjectPreparationGUI} shows the snapshot of the GUI of the tool at the final stage of time-frequency alignment (at 6:41 of the video ``10'' in Fig.~\ref{fig:tutorialVideos}).
In the following sections, we will use video snapshots of this YouTube channel~\cite{worldYouTube} for the introduction. 

\subsubsection{Outline of the morphing object preparation}
The upper right image of Fig.~\ref{fig:mObjectPreparationGUI} shows two time-frequency representations of the filter parameter $\mathbf{sl}$ and 
two waveform representations overlaid with each other respectedly.
These parameter images are half transparent, and it is easy to observe their alignment status.

We put one set of morphing parameters on the ``canonical'' time-frequency axis.
This axis is fixed and does not change throughout the morphing object preparation procedure.
We put the other morphing parameters on the ``non-linear'' time-frequency axis.
The user aims to deform this ``non-linear'' axis to align representations using temporal anchors (vertical lines) and frequency anchors (circles on the lines).
The middle left plot shows the distance between two overlaid images.
The user's task is to minimize the distance by deforming the ``non-linear'' time-frequency axis.
Pursuing this task does not require expertized knowledge and skills. 
We successfully used this video tutorial and a brief tutor introduction in lesson classes.



\subsubsection{Step-by-step procedure}
\begin{figure}[tbp]
\begin{center}
\includegraphics[width=0.8\hsize]{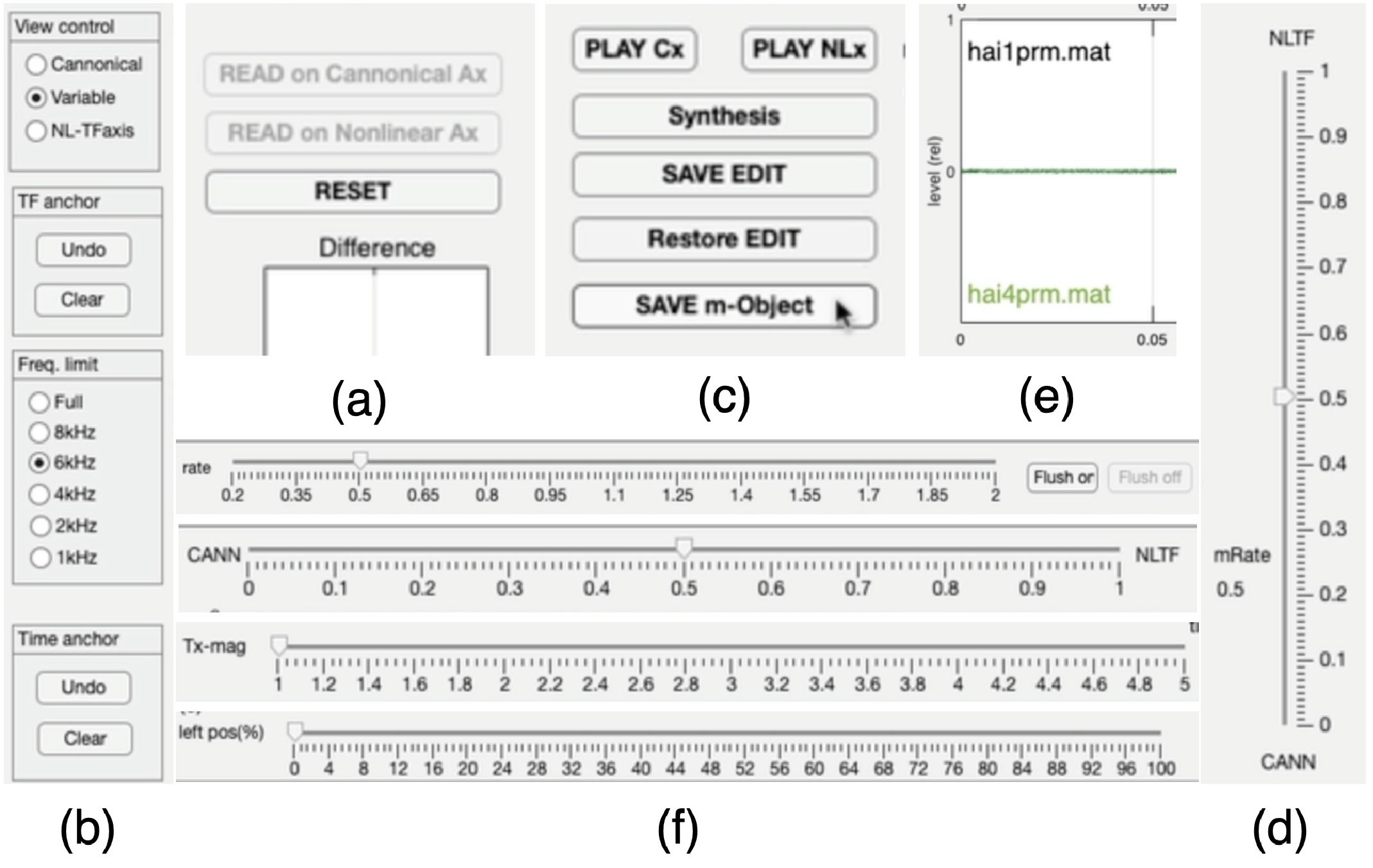}\\
\caption{GUI control parts for manipulation.}
\label{fig:guiParts}
\end{center}
\end{figure}

Figure~\ref{fig:guiParts} shows magnified views of the GUI control parts of the morphing object preparation tool (Fig.~\ref{fig:mObjectPreparationGUI}).
This section introduces the morphing object preparation procedure using the controllers step-by-step.

\paragraph{\underline{Reading VOCODER parameter files:}}
The first step is to read two sets of VOCODER parameters to the ``canonical'' axes and ``non-linear'' axes.
We use buttons on the top left (a) of the GUI.
They are \ovalbox{\textsf{READ on Canonical Ax}},  \ovalbox{\textsf{READ on Nonlinear Ax}}, and  \ovalbox{\textsf{RESET}} from top to bottom.
Clicking the  \ovalbox{\textsf{READ on Canonical Ax}} button invokes file input dialogue to read a VOCODER parameter file.
After reading the file, read the other VOCODER parameter using the \ovalbox{\textsf{READ on Nonlinear Ax}}  button.
Completion of this step fills the spectrogram and the waveform displays.
The image (e), shows the VOCODER file names; 
``canonical'' file name (colored black), and the ``non-linear'' file name (colored green).

The initial condition of the displays is half-transparent, and two overlaid representations are visible.
Selecting the ``\textsf{Canonical}'' button on top of the image (b) only shows parameters on the ``canonical'' axes.
The ``\textsf{NL-TFaxis}'' button only shows the parameters on the ``non-linear'' axes.
The ``\textsf{Variable}'' button resumes. 

\begin{figure}[tbp]
\begin{center}
\includegraphics[width=0.8\hsize]{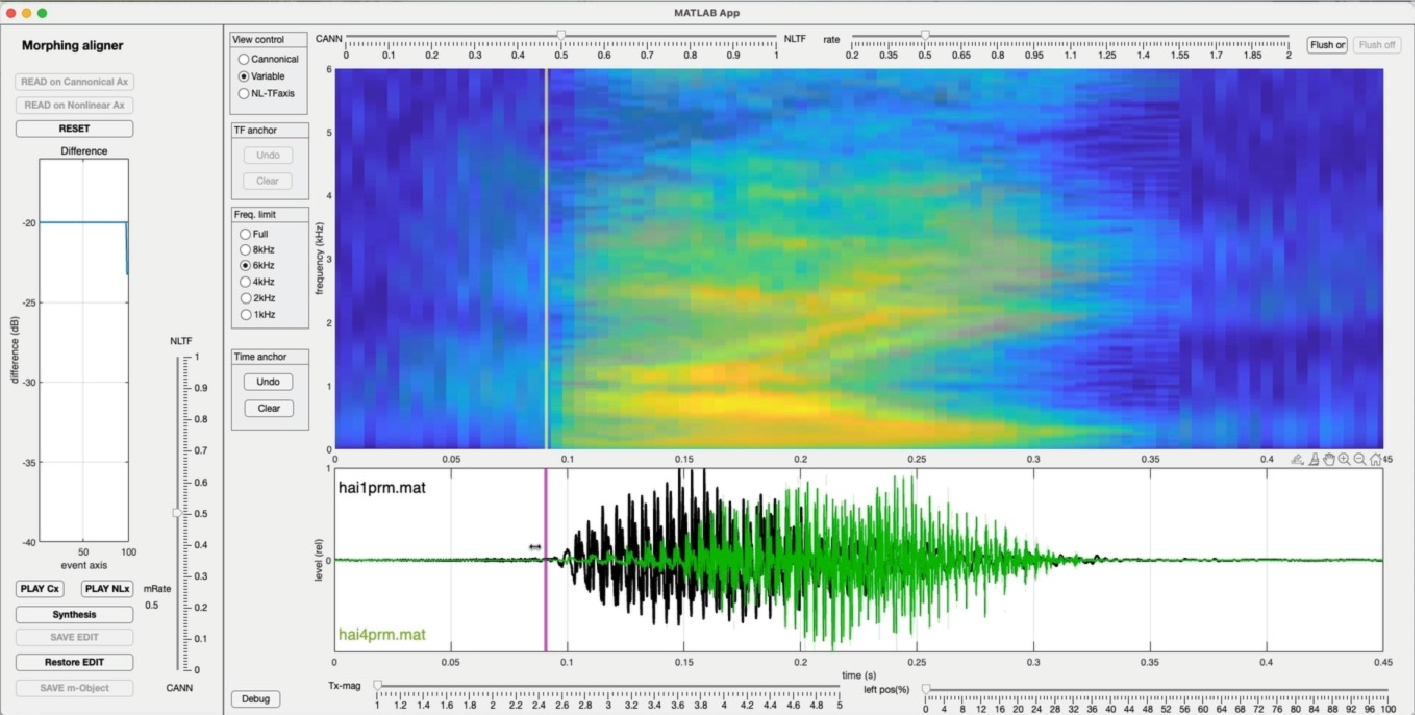}\\
\caption{Temporal anchor assignment and adjustment (at 1:08 in video).}
\label{fig:timeAnchoring}
\end{center}
\end{figure}

\paragraph{\underline{Temporal anchor assignment and adjustment:}}
The next step is to assign and adjust temporal anchors.
Figure~\ref{fig:timeAnchoring} shows just after the first temporal anchor setting and adjustment (at 1:08 in the video).
The alignment of parameters is poor at this point (0.2~s to 0.3~s; the second formant: F2 trajectories are displaced).
The user can check alignment by synthesizing speech from parameters using \ovalbox{\textsf{Synthesis}}, \ovalbox{\textsf{PLAY Cx}}, and \ovalbox{\textsf{PLAY NLx}} buttons.
Click plays back ``morhed,'' ``cannonical,'' and ``non-linear'' parameters' sound.

While the mouse cursor shape is ``arrow'' or a small ``cross-hair,'' clicking in the waveform display places a temporal anchoring point.
It also draws a vertical line in each display.
The \ovalbox{\textsf{Undo}} button at the bottom of the image (b) removes the latest placed temporal anchor.
The \ovalbox{\textsf{Clear}} button removes all temporal anchors. 

When the mouse cursor is close to one of the temporal anchors, its shape changes to a ``hand.''
Then, clicking an anchor (a thin vertical line) changes the anchor to a thick purple line and draggable (Figure~\ref{fig:timeAnchoring}).
Dragging the anchor deforms the green waveform and the distance trajectory on the left middle plot changes.
The green distance trajectory shows a sudden drop at the end.
Adding and adjusting temporal anchors with deformation, the temporal alignment procedure finished at 2:25.
Completing this process assigns the temporal anchors to 
parameters displayed on \underline{\bf ``non-linear''} axis. 

\begin{figure}[tbp]
\begin{center}
\includegraphics[width=0.8\hsize]{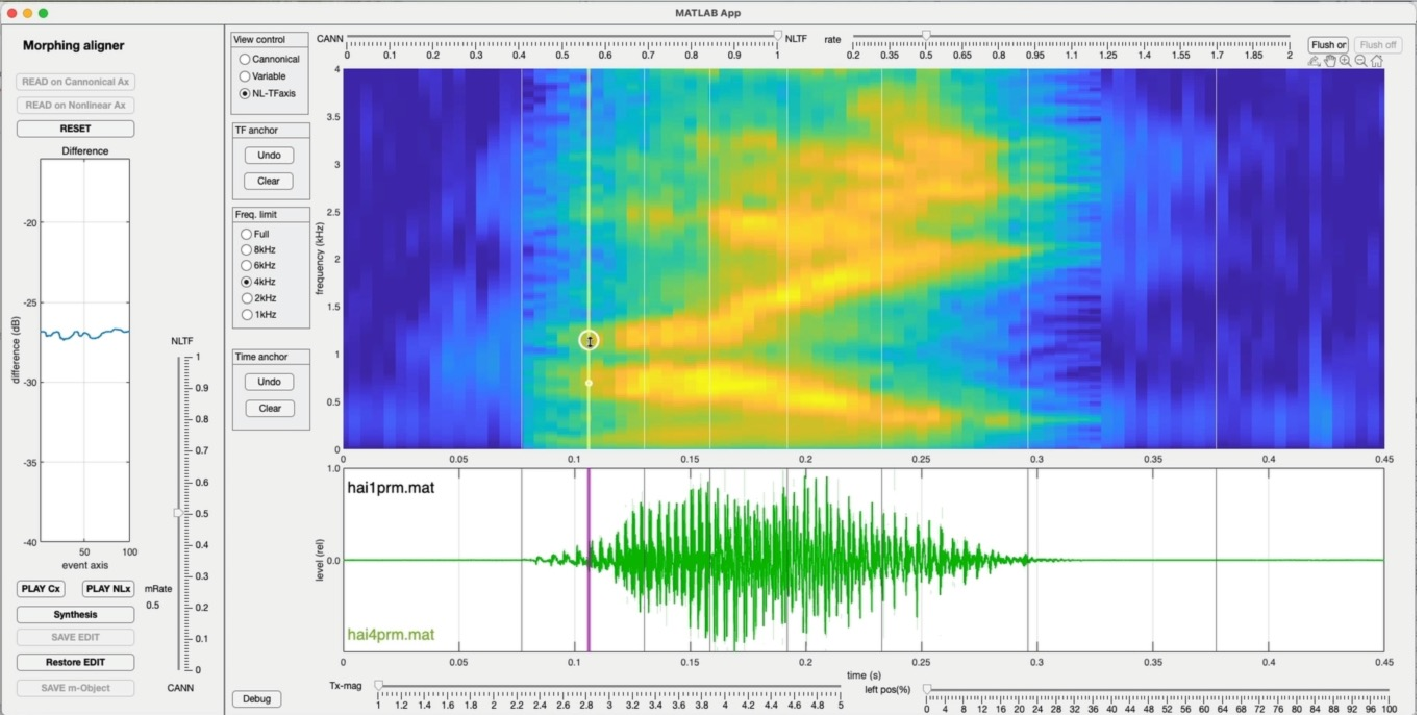}\\
\caption{Frequency anchor assignment on the ``non-linear'' representation. (at 3:09)}
\label{fig:frequencyAnchoring}
\end{center}
\end{figure}

\paragraph{\underline{Frequency anchor assignment:}}
The frequency anchors are attached to the representation on the ``non-linear'' spectrogram axis.
Figure~\ref{fig:frequencyAnchoring} shows after the second frequency anchor assignment (at 3:09).
When nearing one of the temporal anchor lines in the spectrogram display, the mouse cursor shape changes into a small ``cross-hair.''
It indicates that frequency anchor assignment is possible.
In this situation, clicking places a frequency anchor and briefly displays a large yellow circle to confirm the assignment.
Note that the frequency axis is magnified (see ``Freq. limit'' panel on the image (b)). 

Similar to temporal anchors, ``undo'' and ``clear'' functions are available on the panel ``TF anchor'' of (b).
Completing this process assigns the frequency anchors associated with parameters on \underline{\bf ``non-linear''} axis. 

\paragraph{\underline{Frequency anchor alignment:}}
Figure~\ref{fig:frequencyAnchoring} shows adjusting an anchor on the F2 trajectory at 0.16~s.
The ``hand'' shaped mouse cursor indicates that it focuses on the closest adjustable anchor.
Note that the frequency axis is further magnified (the limit is 2~kHz) to make alignment easier.
The displays show time-frequency representation and waveform on the ``canonical'' axes.

\begin{figure}[tbp]
\begin{center}
\includegraphics[width=0.8\hsize]{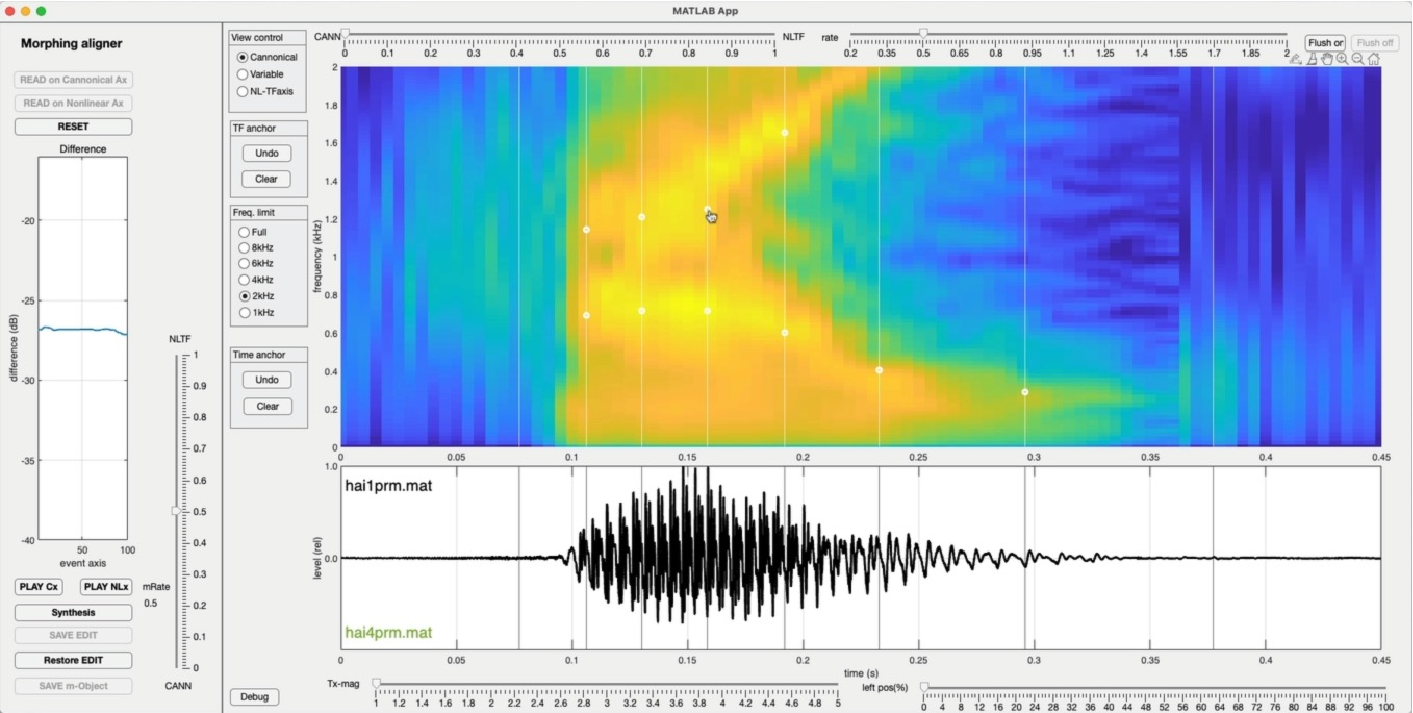}\\
\caption{Frequency anchor alignment on the ``Canonical'' representation. (at 5:04)}
\label{fig:frequencyAlignment}
\end{center}
\end{figure}

The user's task is to adjust the location of the selected anchor (attached on the ``non-linear'' axis) to the relevant location.
In Fig.~\ref{fig:frequencyAlignment}, the user tries to move and adjust the anchor on the F2 trajectory.
Note that the location of the temporal anchors can also be adjusted during this process.
Completing this process assigns the temporal anchors and frequency anchors associated with parameters displayed on the \underline{\bf ``canonical''} axes. 

\paragraph{\underline{Results inspection and saving morphing objects:}}
Using buttons on the lower left of the GUI (image (c) of Fig.~\ref{fig:guiParts}), the user can inspect the synthesized sounds and save the results at any time.
The \ovalbox{\textsf{SAVE EDIT}} button saves the intermediate state of assignment and alignment.
After saving the state, users can terminate the application.
The \ovalbox{\textsf{Restore EDIT}} button recovers the state of assignment and alignment, and users can resume the terminated task.
Moving the morphing rate slider, (d) of Fig.~\ref{fig:guiParts} playbacks the morphed sound on each mouse-up event (from 6:18 to 6:34).
When satisfied with the results, click of \ovalbox{\textsf{SAVE m-Object}} saves the results as re-usable morphing objects (at 6:41).

\paragraph{\underline{Anchoring to new VOCODER parameters:}}
The temporal and frequency anchors can be assigned and aligned to a new VOCODER parameter file using this tool and a morphing object.
This procedure produces a new morphing object.
First, read a new VOCODER parameter file using \ovalbox{\textsf{READ on Canonical Ax}} button.
Then, read a morphing object file using \ovalbox{\textsf{READ on Nonlinear Ax}} button.
This procedure brings the state to the beginning of \underline{\bf ``Frequency anchor alignment.''}
The remaining processes are the same (from 7:42 to 13:51).
Note that the temporal anchor \underline{\bf position} is also adjustable here.

\subsection{Multi-item-morphing demonstration tool}
\begin{figure}[tbp]
\begin{center}
\includegraphics[width=\hsize]{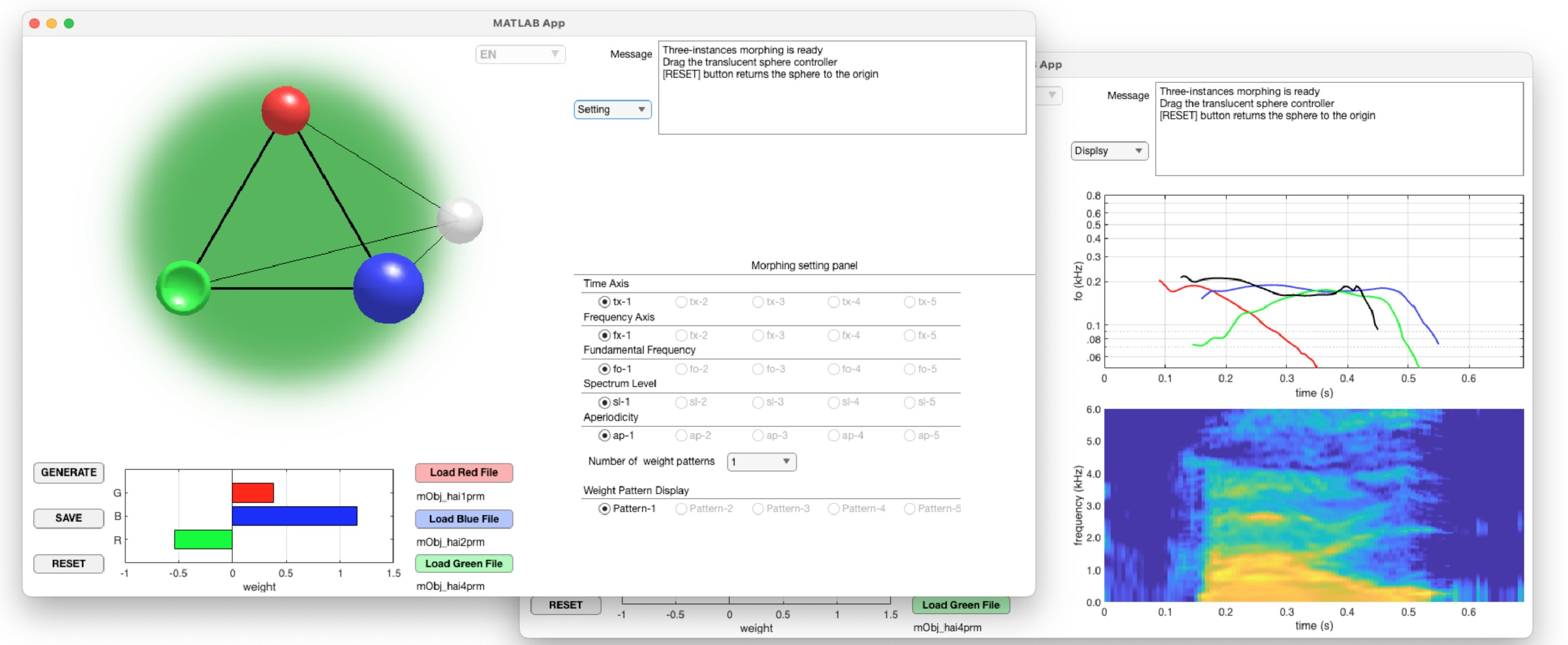}\\
\caption{Application snapshots of three instances morphing demonstration (setting and display modes).}
\label{fig:threeWayGeneralGUI}
\end{center}
\end{figure}

We prepared an interactive application of the generalized voice morphing demonstration using the modern environment ``App Designer'' of MATLAB~\cite{MATLAB} that allows live debugging. 
Figure~\ref{fig:threeWayGeneralGUI} shows a snapshot of the application.
Three colored buttons at the center bottom load three morphing objects on each colored sphere.
After loading morphing objects, the mouse-up event of dragging the half-transparent gray sphere (the control knob) on the left-top image generates a morphed speech sound.
The location of the control knob sets three morphing weights using area coordinate~\cite{Li2011earthSci}.
The weights define the volume of the corresponding color spheres.
A negative valued weight deforms the sphere into a bowl shape.
The bottom right area with many radio buttons enables up to five weighting patterns, making it easy to explore ``parameter-specific morphing''~\cite{Skuk2014JSLHR,Skuk2020JSLHR,Nussbaum2022cognition}.
The area also can visualize the morphed attributes.
An interesting application is to assign divergent speech samples to spheres and explore ``parameter-specific morphing.''

\section{Discussion}
The most crucial issue of this set of tools based on classical VOCODER is its sound quality.
The sound quality of neural VOCODERs is significantly higher than that of classical VOCODERs.
One possible solution is replacing the tool's classical VOCODER with a neural VOCODER that shares similar VOCODER parameters~\cite{Yasuda2021CSL,Yoneyama2023ieee}.
This solution needs a framework to ensure precise control of the acoustic parameters of the generated morphed sounds with the users' intention.

The rapid progress of speech technologies by deep learning makes automatic assessment of speech inevitable~\cite{cooper2024ast}.
While at the same time, researchers' basic understanding of speech science is increasingly important.
Providing lightweight, easy-to-use speech research tools like ours will help with these needs.

Finally, the generalized morphing framework is general enough to replace VOCODER parameters with latent variables of end-to-end speech technologies~\cite{sisman2020ieee,Tan2024ieeePami,cooper2024ast}.
We speculate this leads to a prospective merger.

\section{Conclusion}
We extended our WORLD-based tools~\cite{worldTools2023} to make generalized morphing~\cite{kawahara2013apsipa,kawahara2015book} accessible.
We made the tools open-source~\cite{worldToolPage} and prepared a YouTube Channel~\cite{worldYouTube} and assisting materials.
We hope extended WORLD-based tools help researchers and students acquire deep implicit knowledge of speech production, perception, and signal processing while benefiting from the resurgence of deep learning-based speech technologies.

\section*{Acknowledgement}
We appreciate Toshio Irino, Toshie Matsui, and Nao Hodoshima and their students' feedback and comments on the early versions of the WORLD-based tools and the generalized morphing extensions.
The Japan Society for the Promotion of Science (JSPS) Grants-in-Aid for Scientific Research supported this work. The supporting grant numbers are JP20H00291, JP21K19794, JP21H00497, and JP21H04900.



\end{document}